\magnification=1200  
\hsize=14 true cm  
\vsize=20 true cm 
\baselineskip=3.7ex


\hoffset 1cm


\def\aaa{\alpha} 
\def\bbb{\beta}

\def\fff{\phi} 
 
\def\ggg{\gamma}

\def\ooo{\omega} 
\def\ppp{\pi} 

\def\qqq{\theta}

\def\ttt{\tau}

\def\G{\Gamma}


\def\({\left(}
\def\){\right)}
\def\[{\left[}
\def\]{\right]}

\def\2#1#2{\la#1, #2\ra} 
\def\8{\infty}
\def\={\equiv}

\def\b#1{\bar#1}
\def\c#1{{\cal#1}}

\def\dbb{d\kern-1.5ex-\kern-.4ex} 
\def\db{d\kern-2ex-\kern-.4ex}

\def\h#1{\hat#1}

\def\i1#1{\int_{-\infty}^\infty d#1\,\,} 
\def\ii{^{-1}}
 
\def\intt{\int\!\!\int}
\def\ir{\int_{-\infty}^\infty}

\def\la{\langle\, }
\def\llr{{L^2(\rl)}}

\def\notto{\ /\kern-2.2ex \to }
\def\0#1{(#1)}

\def\pl{\partial}

\def\qq{\qquad} 
\def\qed{\vrule height6pt width3pt depth 0pt}

\def\ra{\,\rangle} 
\def\rat#1#2{{{#1}\over{#2}}}
   
\def\ref{\null}
\def\rl{{\bf R}} 
\def\rr#1{{{\bf R}^#1}}
\def\sh#1{\hglue#1ex}

\def\st#1{{\sl{#1}\/}} 
\def\sv#1{\vglue#1ex}
  
\def\t#1{\tilde#1}

\def\x{\null} 
\def\zz{{\bf Z}}


\def\cl{\centerline} 
\def\CR{commutation relation}

\def\FT{Fourier transform} 
 
\def\ie{i.e., }

\def\n{\noindent}

\def\resp{respectively}

\def\wrt{with respect to}


\newcount\eq\eq=1

\def\e#1{\eqno(#1.\the\eq)\global\advance\eq by 1$$}
\def\en{\eqno(\the\eq)\global\advance\eq by 1$$}

\def\monthname{\ifcase\month\or January\or
   February\or March\or April\or May\or June\or July\or August
   \or September\or October\or November\or December\fi}

\def\har#1{\smash{\mathop{\hbox to .2in{\rightarrowfill}}
      \limits^{#1}}}
     
\def\harr#1{\smash{\mathop{\hbox to .5in{\rightarrowfill}}
      \limits^{#1}}}
     
\def\refs{\bigbreak{\centerline{\bf References}}
\bigskip\frenchspacing\everypar={\hangindent=2em\hangafter
=1}\parindent=0pt\parskip=\smallskipamount}


\cl{(Appeared in \sl Journal of Mathematical Physics \bf
35\rm, 1172-1376, 1994)}

\sv2

\cl{\bf Deformations of Gabor Frames}

\sv2

\cl{\bf Gerald Kaiser}

\cl{Department of Mathematical Sciences}

\cl{University of Massachusetts at Lowell}

\cl{Lowell, MA 01854, USA}

\cl{email: kaiserg@woods.ulowell.edu}

\sv1

\cl{June, 1993}

\sv4

\cl{\bf ABSTRACT}

\sv2

\n The quantum mechanical harmonic oscillator Hamiltonian
$H=(t^2-\pl_t^2)/2$ generates a one--parameter unitary group
$W\0\qqq=e^{i\qqq H}$ in $\llr$ which \st{rotates} 
the time--frequency plane.  In particular, $W(\ppp/2)$ is the
Fourier transform. When $W\0\qqq$ is applied to any frame of
Gabor wavelets, the result is another such frame with identical
frame bounds.  Thus each Gabor frame gives rise to a
one--parameter family of frames, which we call a
\st{deformation} of the original.  For example,  beginning with the
usual tight frame $\c F$ of Gabor wavelets generated by a 
compactly supported window $g\0t$ and parameterized by a
regular lattice in the time--frequency plane,  one obtains a family
$\{\c F_\qqq: 0\le\qqq< 2\ppp\}$ of frames generated by the
\st{non--compactly supported} windows $g\,^\qqq=W\0\qqq g$,
parameterized by rotated versions of the original lattice.  This
gives a method for constructing tight frames of Gabor wavelets for
which neither the window nor its \FT {} have compact support.
When $\qqq=\ppp/2$, $\c F_\qqq$ is the well--known Gabor
frame generated by a window with compactly supported \FT.  The
family $\{\c F_\qqq\}$ therefore interpolates these two familiar
examples.

\sv2

\n PACS numbers:  02.20.+b,  03.65.--w.

\vfill\eject

\n {\bf 1. Introduction}

\sv1

\n If $f\0t$ is a complex--valued, differentiable function, let

$$
(Qf)\0t=tf\0t,\qq (P f)\0t=-if'\0t.
\en 
$Q$ and $P$ extend to unbounded,  self--adjoint  operators on
$\llr$  which satisfy the (Heisenberg) commutation relation $[Q,
P]\=QP-PQ=iI$, where $I$ is the identity operator.    Being
self--adjoint, $Q$ and $P$ generate one--parameter unitary groups
operating on $\llr$,

$$
U\0\ooo=e^{i\ooo Q},\qq V\0s=e^{-isP}.
\en
$U\0\ooo$ and $V\0s$ act by modulation and translation, \resp:

$$
(U\0\ooo f)\0t=e^{i\ooo t}\,f\0t,\qq (V\0sf)\0t=f(t-s).
\en
The local (Lie--algebraic) \CR{} $[Q,P]=iI$ has the global
(Lie--group) equivalent

$$
U\0\ooo\, V\0s=e^{i\ooo s} \,V\0s \,U\0\ooo.
\en
Hence the set of operators $G(\fff, \ooo, s)=e^{i\fff}\,U\0\ooo\,
V\0s$ satisfies the relation

$$
G(\fff, \ooo, s)\,G(\fff', \ooo', s')
=G(\fff+\fff'-\ooo's, \ooo+\ooo', s+s'),
\en
so it forms a group of unitary operators on $\llr$, known as (a 
representation of) the \st{Weyl--Heisenberg group.} The
Gabor transform (or windowed \FT)  may be viewed entirely
in terms of this group.  (See Daubechies \x[1] or Kaiser
\x[2] for general background on the windowed Fourier transform.) 
Namely, given a window function $g\0t$ in $\llr$ with $\|g\|=1$,
define

$$
g_{\ooo, s}\=U\0\ooo\, V\0s\,g.
\en
Then \x(3) shows that $g_{\ooo, s}\0t=e^{i\ooo t}\,g(t-s)$,
which gives the translated and modulated windows that form the
basis of the continuous Gabor transform and its inverse:

$$\eqalign{
\t f(\ooo, s)
&\=\la g_{\ooo, s}\,, f\ra=\i1t e^{-i\ooo t}\,\b g(t-s)\,f\0t, \cr
f\0t &=
\rat 1{2\ppp}\intt_{\rr2} d\ooo\,ds\ g_{\ooo, s}\0t\,
\t f(\ooo, s).
\cr}\en

\sv2

Under certain conditions, a discrete subset of such ``Gabor
wavelets'' $g_{\ooo, s}$ is sufficient to reconstruct $f\0t$ from $\t
f(\ooo, s)$.  For example, suppose that $g$ has compact support in
$[0, \ttt]$ and choose an interval $0<T\le\ttt$. Suppose that the
periodic function

$$
H\0t\=\ttt\sum_{n\in\zz}|g(t-nT)|^2
\en
is bounded above and below by positive constants, \ie

$$
0<A\le H\0t\le B.
\en
Then it can be shown (cf. Daubechies et al.~\x[3], Kaiser \x[4]) that

$$
f\0t=H\0t\ii\sum_{n\in\zz}\sum_{m\in\zz} g_{\ooo_m, s_n}\0t\,
\t f(\ooo_m, s_n),
\en
where $\ooo_m=2\ppp m/\ttt$ and $s_n=nT$.  In fact, \x(9) is
equivalent to

$$
A\|f\|^2\le
\sum_{n\in\zz}\sum_{m\in\zz} |\la g_{\ooo_m, s_n}\,, f\ra|^2
\le B\|f\|^2,
\en
which is just the condition that the set of vectors 
$\{g_{\ooo_m, s_n}: m, n\in \zz\}$ form a \st{frame} in $\llr$ with
frame bounds $A,B$.  This frame can be made tight if we replace
$g$ by the new window

$$
h\0t\=H\0t^{-1/2}\,g\0t.
\en
Since $h$ is also compactly supported, and 
$\ttt\sum_{n\in\zz}|h(t-nT)|^2\=1$, the above shows that 
$\{h_{\ooo_m, s_n}\}$ form a tight frame with bounds $A'=B'=1$,
i.e., a \st{resolution of unity.} Note, however, that 
$\|h\|^2=T/\ttt\le 1$, so this frame is  an orthonormal
basis if only if $T=\ttt$. But if $T=\ttt$, then $g$  (and therefore
also $h$) must be discontinuous in order to satisfy (9); this means
that $\h g\0\ooo$ and $\h h\0\ooo$ have slow decay, giving $\t
f(\ooo, s)$ poor frequency resolution.  (This is a special case of the
\st{Balian-Low theorem;} see Daubechies \x[1].)

\sv2

The above construction depends crucially on the fact that $g$ has
compact support, since it expands $\b g(t-s)\,f\0t$ in a Fourier
series. A similar construction can be made starting with a window
which is compactly supported in frequency (expand in
a Fourier series in the frequency domain). This leads  again to a
resolution of unity in terms of Gabor wavelets.  For windows that
are neither compactly supported in time nor in frequency, both
constructions fail. For such windows, it is generally  
difficult to give explicit constructions of resolutions of unity.  The
theorem proved in the next section provides a method for
generating a one--parameter \st{family} of Gabor frames starting
from an arbitrary Gabor frame.  If the original frame is tight or
orthonormal, then so is each member of the family it generates.    A
general member of this family will have a window which is neither
compactly supported in time nor in frequency, even when the
original window has compact support.

\sv4

\n{\bf 2. Deformations of Frames}

\sv1
\n Consider  the harmonic oscillator Hamiltonian

$$
H\=\rat12(Q^2+P^2)=\rat12(t^2-\pl_t^2),
\en
which is an unbounded and self--adjoint operator on $\llr$. 
Note that

$$
[H, Q]=-iP,\qq [H, P]=iQ.
\en
 Let $W(\qqq)\=e^{i\qqq H}$ be the one--parameter group of
unitary operators generated by $H$, and define the one--parameter
families of operators

$$
Q\0\qqq\=W(\qqq)\,Q\,W(-\qqq),
\qq P\0\qqq\=W(\qqq)\,P\,W(-\qqq).
\en
Then 

$$\eqalign{
\rat d{d\qqq} Q\0\qqq &=i\,[H, Q\0\qqq]
=i\,W\0\qqq\,[H, Q] \,W(-\qqq)=P(\qqq), \cr
\rat d{d\qqq} P\0\qqq&=i\,[H, P\0\qqq]
=i\,W\0\qqq\,[H, P] \,W(-\qqq)=-Q(\qqq).
\cr}\en
These operator equations (with the initial conditions $Q\00=Q,
\,P\00=P$) can be integrated to give

$$
Q\0\qqq=Q\cos\qqq+P\sin\qqq,\qq
P\0\qqq=P\cos\qqq-Q\sin\qqq.
\en
Hence, $W\0\qqq$ ``rotates'' the operators $Q$ and $P$. 
$W\0\qqq$ acts on the  ``global'' operators $U\0\ooo$ and $V\0s$
by

$$\eqalign{
W\0\qqq\,U\0\ooo&=W\0\qqq \,e^{i\ooo Q}
=e^{i\ooo Q(\qqq)}\,W\0\qqq,\cr
W\0\qqq\,V\0s&=W\0\qqq \,e^{-isP}
=e^{-isP(\qqq)}\,W\0\qqq.
\cr}\en
From \x(17) and \x(18), we can compute the action of  $W\0\qqq$
 on the Gabor wavelets $g_{\ooo, s}$:

$$
W\0\qqq\,g_{\ooo, s}=W\0\qqq\,\,e^{i\ooo Q}\,e^{-isP} \,g=
e^{i\ooo Q\0\qqq}\,e^{-is P\0\qqq}\,W\0\qqq\,g.
\en
To obtain an explicit expression, let $g\,^\qqq\=W\0\qqq\,g$. This
is a new window function which  generate a new family of
Gabor wavelets $g\,^\qqq_{\ooo, s}\=U\0\ooo\,V\0s\,g\,^\qqq$.  

\proclaim Theorem 1.  $W\0\qqq$ acts on the Gabor wavelets
$g_{\ooo, s}$ by replacing the window $g$ with $g\,^\qqq$,
rotating the labels $(\ooo, s)$, and multiplying by a phase factor
$\ggg$: 
$$
W\0\qqq\,g_{\ooo, s}
=\ggg(\ooo, s, \qqq)\, g\,^\qqq_{\ooo(\qqq),
s(\qqq)}\,, \en
where 
$$
\ggg(\ooo, s,\qqq)
=\exp\[\rat i4(\ooo^2-s^2)\sin (2\qqq)+i\ooo s \sin^2\qqq\]
\en and
 $$
s\0\qqq=s\cos \qqq+\ooo\sin\qqq, 
\qq\ooo\0\qqq=\ooo\cos\qqq-s\sin\qqq.
\en
The new window $g\,^\qqq$ is given by
$$
g\,^\qqq\0u=(-2\ppp i\,\sin\qqq)^{1/2}
\ir dt\,\exp\[iut\csc\qqq-i(u^2+t^2)\cot\qqq\] g\0t
\en
for $\qqq\ne n\ppp$.   When $\qqq=\ppp/2$, \x(23)
reduces to $g^{\ppp/2}\0u=e^{i\ppp/4}\,\h g\0u$, where $\h g$ is
the \FT{} of $g$.  The singular cases are given by
$g^{n\ppp}\0t=i^n\,g((-1)^n\,t)$, $n\in\zz$.

\sv2

\n\st{Proof:} We need the identity

$$
e^{i(aQ+bP)}=e^{iab/2}\,e^{iaQ}\,e^{ibP},\qq a, b\in\rl\,,  
\en
which is closely related to \x(4).  Setting $\aaa=\cos\qqq$ and
$\bbb=\sin\qqq$, we have

$$\eqalign{
e^{i\ooo Q\0\qqq}&=e^{i\ooo\aaa Q+i\ooo\bbb P}
=\exp\[i\ooo^2\aaa\bbb/2\]e^{i\ooo\aaa Q}\,e^{i\ooo\bbb P},\cr
e^{-isP\0\qqq}&=e^{is\bbb Q-is\aaa P}
=\exp\[-is^2\aaa\bbb/2\]e^{is\bbb Q}\,e^{-is\aaa P}.
\cr}\en
Substituting this into \x(19) and using 

$$
e^{i\ooo\bbb P}\,e^{is\bbb Q}=V(-\ooo\bbb)\,U(s\bbb)
=\exp\[i\ooo s\bbb^2\]e^{is\bbb Q}\,e^{i\ooo\bbb P},
\en
which follows from \x(4), we obtain

$$
W\0\qqq\,g_{\ooo, s}
=\exp\[i(\ooo^2-s^2)\aaa\bbb/2+i\ooo s\bbb^2\]
e^{i(\aaa\ooo+\bbb s)Q}\,e^{-i(\aaa s-\bbb \ooo)P}\,g\,^\qqq,
\en
which is \x(20). Eq.~\x(23) follows from Mehler's formula for the
kernel of $W(\qqq)$  (cf.~Merzbacher \x[5], p.159, where
$\qqq=-t$).  That $g^{\ppp/2}=e^{i\ppp/4}\h g$ follows from
\x(23) by taking the limit $\qqq\to\ppp/2$.  The expression for
$g^{n\ppp}$ then follows from $W(n\ppp)=(W(\ppp/2))^{2n}$ and
${\hat{\hat g}}(t)=g(-t)$. \sh2\qed

\vfill\eject

\proclaim Corollary 2.  Let $\c F=\{g_{\ooo, s}:  (\ooo, s)\in\G\}$ be
any frame of Gabor wavelets, labeled by a discrete set
$\G\subset\rr2$, with window $g$ and frame bounds $0<A\le
B<\8$.   For every real $\qqq$, let   $\G_\qqq=\{(\ooo\0\qqq,
s\0\qqq): (\ooo, s)\in\G\}$ be the rotated version of $\G$ and let
$g\,^\qqq=W\0\qqq\, g$.
 Then the set of vectors   $\c F_\qqq\=\{g\,^\qqq_{\ooo, s}:  (\ooo,
s)\in\G_\qqq\}$ forms a Gabor frame with window $g\,^\qqq$ and
identical frame bounds $A, B$.

\sv2

\n\st{Proof:}  For any $(\ooo, s)\in\rr2$, Theorem 1 gives

$$
g\,^\qqq_{\ooo\0\qqq, s\0\qqq}
=\ggg(\ooo, s,\qqq)\ii W\0\qqq\,g_{\ooo,s}\,.
\en
Hence the Gabor transform \wrt{} the vectors in $\c F_\qqq$ is

$$\eqalign{
\la g\,^\qqq_{\ooo\0\qqq,  s\0\qqq}\,, f\ra
&=\ggg(\ooo, s,\qqq)\,\la W\0\qqq g_{\ooo, s}\,, f\ra  
=\ggg(\ooo, s,\qqq)\,\la g_{\ooo, s}\, , W(-\qqq)\,f\ra \cr
&=\ggg(\ooo, s,\qqq)\,\la g_{\ooo, s}\, , f\,^\qqq\ra
\cr}\en
by the unitarity of $W(\qqq)$, where 
$f\,^\qqq\=W(-\qqq)f\in\llr$.  Thus

$$
\sum_{(\ooo, s)\in\G_\qqq} | \la g\,^\qqq_{\ooo, s}\,, f\ra|^2=
 \sum_{(\ooo, s)\in\G} | \la g\,^\qqq_{\ooo\0\qqq, s\0\qqq}\,,
f\ra|^2 =\sum_{(\ooo, s)\in\G} | \la g_{\ooo, s}\, , f\,^\qqq\ra|^2.
\en
Since $\c F$ is a frame, 

$$
A\|f\,^\qqq\|^2\le
\sum_{(\ooo, s)\in\G} | \la g_{\ooo, s}\, , f\,^\qqq\ra|^2
\le B\|f\,^\qqq\|^2.
\en
By the unitarity of $W(-\qqq)$,  $\|f\,^\qqq\|^2=\|f\|^2$.  Hence
\x(30) shows that $\c F_\qqq$ is a frame as well, with the same
frame bounds $A$ and $B$.  \sh2\qed

\sv2

For example, if $\c F$ is the frame generated by a
compactly supported window $g$ as in \x(10), then
$\c F_\qqq$ is generated by the window
$g\,^\qqq$.  This gives a one--parameter family of frames $\c
F_\qqq$, where only $\c F_{n\ppp}$ have compactly supported
windows ($n\in\zz$).  However, the frames $\c F_\qqq$ with
$\qqq=(n+\rat12)\ppp$ have windows which are compactly
supported in the frequency domain.   The family $\{\c F_\qqq\}$
therefore interpolates the  two familiar types of easily constructed
Gabor frames having windows with compact supports in time and
in frequency, \resp.  We call the family $\c F_\qqq$ a
\st{deformation} of the frame $\c F$.  Note that since
$g^{2\ppp}\0t=-g\0t$, the frame $\c F_{2\ppp}$ is essentially
equivalent to $\c F_0=\c F$, and  we only get distinct frames for
$0\le\qqq<2\ppp$.

\sv2

After this note was completed, I learned that work along similar
(but inequivalent) lines has been done by R.~G.~Baraniuk and
D.~L.~Jones \x[6].  They consider deformations of Gabor frames by
operators that  \st{shear} or \st{chirp} the time--frequency plane. 
I thank Bruno Torresani for pointing this work out to me.

\sv6

\refs

[1]  I.~Daubechies (1992), \sl Ten Lectures on Wavelets, \rm
SIAM, Philadelphia.

[2] G.~Kaiser (1994), \sl A Friendly Guide to Wavelets, \rm
Birkh\"auser, Boston.

[3]  I.~Daubechies, A.~Grossmann and Y.~Meyer (1986), \sl Painless
non--ortho\-gonal expansions, \rm J. Math. Phys. {\bf 27},
1271-1283.

[4] G.~Kaiser (1984), \sl A sampling theorem in the joint
time--frequency domain, \rm University of Lowell preprint.

[5] E.~Merzbacher (1961), \sl Quantum Mechanics, \rm Wiley, New
York.

[6]  R.~G.~Baraniuk and D.~L.~Jones (March 1993 preprint), \sl New
orthonormal bases and frames using chirp functions, \rm  to appear
in IEEE Transactions on Signal Processing, special issue on Wavelets
in Signal Processing.

\bye